# X-ray Spectrum of the Black Hole Candidate X1755-338

H.C. Pan[1,3], G. K. Skinner[1], R. A. Sunyaev[2], and K. N. Borozdin[2]
[1] *School of Physics and Space Research, University of Birmingham, Edgbaston, Birmingham B15 2TT*
[2] *IKI, Space Research Institute, Profsoyuznaya, 84/32, Moscow 117296, Russia*
[3] *Department of Physics (Theoretical Physics), University of Oxford, 1 Keble Road, Oxford OX1 3NP*

**ABSTRACT**
We report the first detection of a hard power-law tail in the X-ray spectrum of the black hole candidate (BHC) binary X1755-338, which was observed in 1989 March-September during the *TTM* Galactic Centre survey. In addition, an ultrasoft thermal component with a temperature of $\sim 1.1 - 1.4$ keV was also detected. We demonstrate that the soft and hard X-ray components of X1755-338 vary independently, as in the spectra of the well known BHCs LMC X-1, LMC X-3 and GX339-4 in their high (intensity) state. If the hard tail observed from X1755-338 is generated near the black hole by energetic electrons up-scattering low energy photons, the un-correlated variations imply that the soft X-rays from accretion disc may not be the main photon seeds needed for inverse Comptonization. The *TTM* observations strongly suggest that X1755-338 does indeed belong to the family of BHCs.

## 1 INTRODUCTION

More than two dozen X-ray binary systems, which have been classed as 'Black Hole Candidates' (e.g. Tanaka 1989; Stella et al. 1994), have now been detected in our galaxy and the Large Magellanic Cloud. In some cases the evidence for the presence of a black hole comes directly from dynamical studies, e.g. Cyg X-1 (Paczynski 1974), LMC X-3 (Cowley et al. 1983), LMC X-1 (Hutchings et al. 1987), A0620-00 (McClintock & Remillard 1986), GS2023+338 (Casares et al. 1992), and GS1124-68 (GRS1124-68) (Remillard et al. 1992). In others the evidence for the presence of a black hole is less direct; X-ray light curves and spectra indicate that the systems are all members of a single class and show characteristics which are quite distinct from those of X-ray binaries known to contain a neutron star (e.g. Tanaka 1989; 1990).

The X-ray spectra typically have a flat power-law high energy tail. In addition there is frequently an ultrasoft thermal component. Some sources are observed in two states. Among the best established black hole binaries a 'high' state with a composite spectrum including an ultrasoft component and a flat power-law type tail at high energies, is shown by Cyg X-1, LMC X-3, LMC X-1, A0620-00 and GS1124-68 (Tanaka 1989; Ebisawa et al. 1994), while a low state in which the X-ray spectrum can be approximated by a power-law with photon index $\sim 1.5$-$1.8$ as shown by Cyg X-1 (e.g. Tanaka 1989; 1990), GS2023+338 (in't Zand et al. 1992; Pan et al. 1993), and GS1124-68 (Ebisawa et al. 1994). When BHC sources are in the 'high' state, the soft and hard components vary independently (e.g. Tanaka 1989; 1990).

X1755-338 is an X-ray binary source located in the general direction of the Galactic Centre. It was previously noted as an unusually soft X-ray source by Jones (1977) and was suggested later as a black hole candidate by White & Marshall (1984), and White et al. (1984), based on its similar location to a group of BHCs on the X-ray color-color diagram. The *Einstein* SSS observations of X1755-338 (White & Marshall 1984) give an interstellar hydrogen column density $N_\mathrm{H} = (2.2 \pm 1.2) \times 10^{21}$ H cm$^{-2}$, far below that expected from sources close to the Galactic Bulge. X-ray dips with 4.4 hour period were observed from X1755-338 with *EXOSAT* (White et al. 1984; Mason et al. 1985). The X-ray flux in the 1-9 keV band decreases by as much as $\sim 40\%$ in the dips which appear to be energy independent as suggested by insignificant change of the hardness ratio from non-dip to dip periods (White et al. 1984).

We observed X1755-338 in 1989 March-September during the period of the *TTM* Galactic Centre Survey. In this letter we present results of the *TTM* observations of X1755-338 and report the first detection of a power-law hard tail in the spectrum of X1755-338.

## 2 OBSERVATIONS

The *TTM* is a coded mask imaging spectrometer on board the *KVANT* module of the *MIR* space station. It has an

Table 1. Spectral parameters of X1755-338, GX339-4, LMC X-1 & LMC X-3.

| Date | Disc-blackbody[1] | | Power-law[2] | | $N_{\rm H}$ [3] | $L_{\rm DB}^{[4]} + L_{\rm PL}^{[4]}$ | $\chi^2/dof$ |
|---|---|---|---|---|---|---|---|
| | $r_{\rm in}(\cos i)^{1/2}$ | $T_{\rm in}$ | Scale | $\alpha$ | | | |
| X1755-338[5] | | | | | | | |
| 89 Mar 31 | $5.2 \pm 2.0$ | $1.14 \pm 0.22$ | $1.37 \pm 0.43$ | $1.9 \pm 0.9$ | $6 \pm 12$ | $0.6 + 0.6$ | 24/25 |
| 89 Aug 23 | $5.2 \pm 0.7$ | $1.33 \pm 0.06$ | $0.88 \pm 0.25$ | $2.7 \pm 0.4$ | $< 1$ | $1.3 + 0.5$ | 30/24 |
| 89 Sep 10 | $5.1 \pm 1.7$ | $1.28 \pm 0.15$ | $2.33 \pm 0.56$ | $2.3 \pm 0.6$ | $< 3$ | $1.0 + 1.1$ | 17/24 |
| GX339-4[5] | | | | | | | |
| 89 Mar 25 | $18.1 \pm 0.4$ | $0.78 \pm 0.01$ | $0.59 \pm 0.08$ | $2.1 \pm 0.3$ | $< 0.1$ | $1.0 + 0.1$ | 69/25 |
| LMC X-1[5] | | | | | | | |
| 88 Nov 13-Dec 2 | $39 \pm 10$ | $0.83 \pm 0.06$ | $0.25 \pm 0.06$ | $2.7 \pm 1.1$ | $< 5$ | $7.0 + 4.0$ | 21/25 |
| 88 Dec 4-5 | $36 \pm 5$ | $0.90 \pm 0.07$ | $0.72 \pm 0.10$ | $1.2 \pm 0.4$ | $< 2$ | $9.0 + 9.0$ | 22/25 |
| LMC X-3[5] | | | | | | | |
| 89 Mar 2 | $28 \pm 4$ | $1.16 \pm 0.07$ | $0.85 \pm 0.03$ | $2.3 \pm 0.2$ | $< 1$ | $19.0 + 12.0$ | 30/23 |
| 89 May 30-Jun 10 | $24 \pm 5$ | $0.94 \pm 0.05$ | $1.06 \pm 0.05$ | $2.1 \pm 0.2$ | $< 2$ | $5.0 + 14.0$ | 29/25 |

Notes:
(1) $r_{\rm in}$ is the radius of the inner accretion disc in km and $i$ the system inclination angle; $T_{\rm in}$ is the blackbody temperature in keV at $r_{\rm in}$. See Makishima et al. (1986) for details.
(2) $\alpha$ is the photon index and Scale in units of $10^{-3}$ photons sec$^{-1}$ cm$^{-2}$ keV$^{-1}$ at 10 keV.
(3) $N_{\rm H}$ is the equivalent hydrogen column density in units of $10^{21}$ cm$^{-2}$.
(4) $L_{\rm DB}$ and $L_{\rm PL}$ are the 2-30 keV luminosities of the disc-blackbody and power-law components in units of $10^{37}$ erg sec$^{-1}$. Corrections are applied to take account of low energy absorption.
(5) Distance to LMC X-1 and LMC X-3: 50 kpc (Stothers 1983); to GX339-4: 4 kpc (Doxsey et al. 1979); and to X1755-338: $< 9$ kpc (Mason et al. 1985).

8 degree square full-width at half-maximum (FWHM) field of view (FOV) and is capable of producing images with 2 arcmin angular resolution in the 2-30 keV energy range. Its energy resolution is about 18% at 6 keV (FWHM) across the whole FOV. The instrumental details are given in Brinkman et al. (1985).

In the period of the *TTM* Galactic Centre survey, we observed X1755-338 on 1989 March 20, 31, April 1, August 22-23, and September 9-10. During the observations, the source intensity varies in the range of 35-105 mCrab (2-10 keV). The X-ray intensity in 1989 August-September period appears to be nearly twice that observed in 1989 March-April. The large data gaps in the *TTM* light curves prevent us from definitely identifying any X-ray dips.

## 3  X-RAY SPECTRUM

The X-ray spectra of X1755-338 have been extracted from images formed by a correlation technique in 31 energy channels covering the energy range 2-30 keV. In order to achieve better statistics, we have studied the X-ray spectrum accumulated during each observational day. We find that the source spectrum consists of an ultrasoft thermal component and a hard tail above $\sim 6 - 10$ keV, which is similar to the high state spectrum of the BHCs LMC X-1, LMC X-3 and GX339-4, also observed with the *TTM*. To our knowledge, this is the first time that a hard tail has been observed from X1755-338.

We have modelled the hard tail with a power-law of photon index $\sim 2$, and the ultrasoft thermal component with several models; a blackbody; a multi-temperature disc blackbody (Mitsuda et al. 1984; Makishima et al. 1986); unsaturated Comptonization (White et al. 1985); Sunyaev-Titarchuk (1980) Comptonization. All of these models give adequate descriptions to our data and we cannot distinguish a 'best' model with the statistics available. Treves et al. (1990) have argued that the unsaturated Comptonization model cannot apply in the case of LMC X-3 since the parameters obtained in fitting the *GINGA* spectra of this source are not mutually consistent. On the other hand a single blackbody model is unlikely to be realistic. For the present work we have chosen to present the data in terms of a model comprising a multi-temperature disc blackbody and a power-law. Low energy absorption is considered, using the model of Morrison & McCammon (1983). We list in Table 1 the derived spectral parameters for the X1755-338 spectra obtained on 1989 March 31, August 23, and September 10, together with those of LMC X-1, LMC X-3, and GX339-4. The errors quoted in Table 1 are at the 90% confidence level.

We plot in Fig. 1 three spectra of X1755-338 (see Table 1). In each case a significant hard tail is seen in the 15-30 keV region. Comparing two spectra on August 23 and September 10 we note that while the X-ray flux below about 8 keV remains almost unchanged, the flux above 8 keV significantly increases from August 23 to September 10. The flux in the range 8-30 keV in the spectrum of March 31 is between those observed on August 23 and September 9. However, the flux below 8 keV on March 31 is almost a factor of two lower than in the other two spectra.

For comparison, we also plot in Fig. 1 the spectra of LMC X-1, LMC X-3 and GX339-4. Similar independent variations of the two spectral components are also seen in the case of LMC X-1 and LMC X-3. While the low energy X-ray intensity of LMC X-1 changes very little between 1988 November 13-December 2 and December 4-5, there is a conspicuous increase in high energy band tail on December 4-5. On the other hand, while the hard tail of LMC X-3 shows little change between the observations made on 1989 March 2 and May 31-June 10, the fluxes below 8 keV differ by more than a factor of two. It is clear from Table 1 that the soft X-ray variation of X1755-338, LMC X-1 and LMC X-3 is due mainly to the change in the temperature of the soft X-ray emission region. When the temperature increases, the source become brighter.

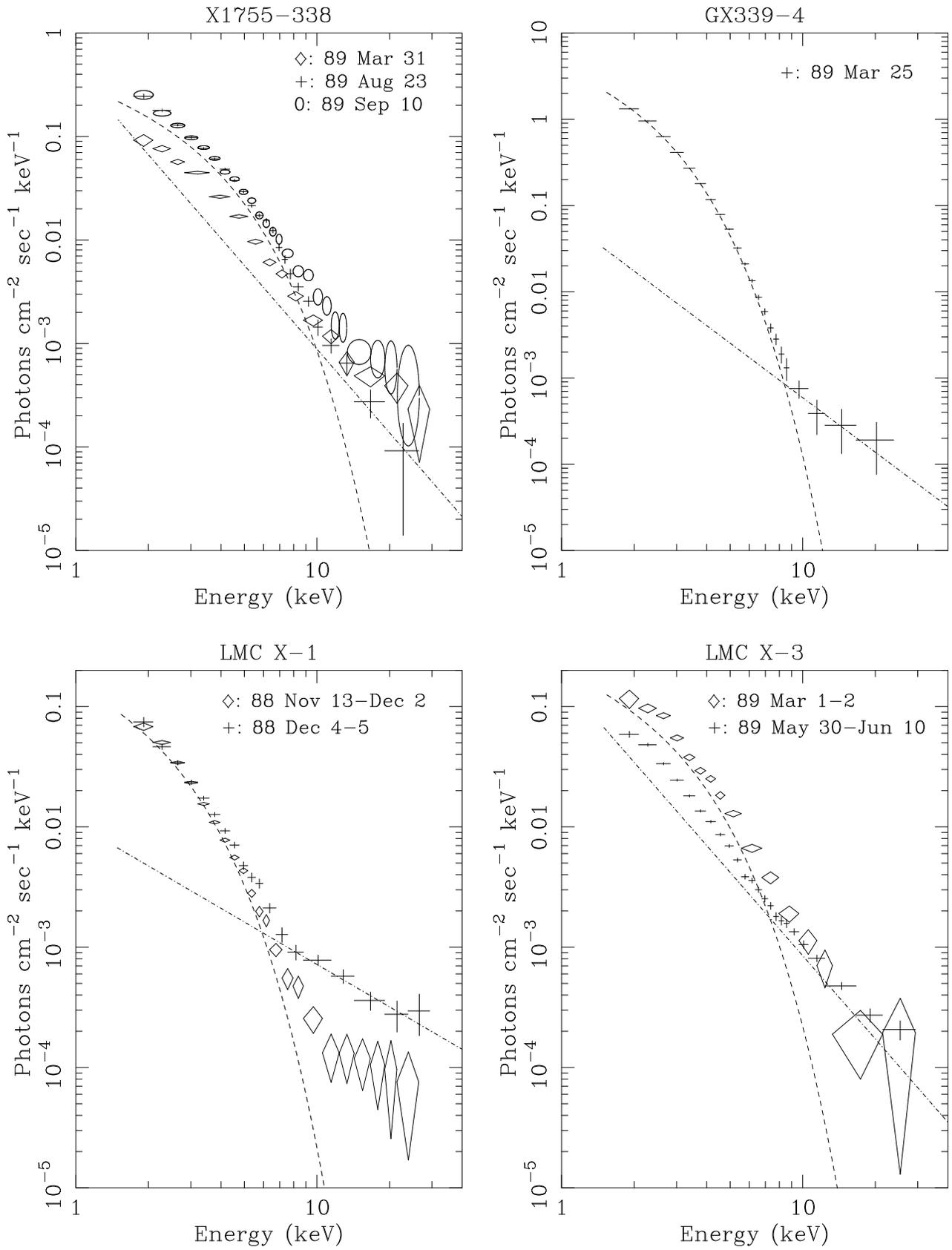

**Figure 1.** X-ray spectra of X1755-338, GX339-4, LMC X-1, and LMC X-3 from *TTM* observations. The dash line represents the disc blackbody component and the dash-dot line is the power-law tail.

## 4 DISCUSSION

The *KVANT/MIR* observations of the BHCs Cyg X-1, GS2023+338, GS2000+25, 1E1740.7-2942 and GX339-4 show that the hard power-law emission extends up to 130-500 keV (Sunyaev et al. 1992; Skinner et al. 1991; Dobereiner et al. 1989), which indicates that the hard power-law emissions of these sources must come from a very high temperature region, presumably close to the black hole, which is hot and transparent. In the 'two-temperature' disc model by Shapiro et al. (1976), soft photons, both internal and external, are Compton up-scattered in a hot plasma ($\sim 10^9$ K) to generate the hard X-ray emission. Our observations reveal no correlation between soft and hard X-ray components, as in the case of other BHCs in their high state. The un-correlated variations may suggest that the external soft photons from the accretion disc are not the main seeds for the inverse Comptonization because the change in the number of photons injected from the accretion disc as the soft component varies would then make the Comptonized component vary accordingly.

Although the spectra of X1755-338, LMC X-1, LMC X-3 and GX339-4 look similar, there are significant differences among their spectral parameters. The temperature of the soft component of the spectra of X1755-338 is significantly higher than other sources (see Table 1). The parameter $r_{in}(\cos i)^{1/2}$ of the disc blackbody, which is derived from the disc blackbody normalization, taking into account the source distance, is also different. The mean values of $r_{in}(\cos i)^{1/2}$ found from fitting the spectra of LMC X-1, LMC X-3 and GX339-4 are 37, 26 and 18 km respectively, all greater than the value of 9 km, which is three times Schwarzschild radius of a black hole with solar mass. However for X1755-338 the value of $r_{in}(\cos i)^{1/2}$ is only about 5 km. If the compact object of X1755-338 is indeed a black hole and we take 9 km as the lower limit of the radius of the inner accretion disc $r_{in}$, then the lower limit of the inclination angle $i$ can be derived and is $i > 72°$.

In summary, the *TTM* observations of X1755-338 show that the spectrum of X1755-338 can be fitted with an ultrasoft thermal component and a hard tail above $\sim 6 - 10$ keV, which is similar to the tails found in the well known BHCs LMC X-1, LMC X-3 and GX339-4. The observations strongly suggest that X1755-338 does indeed belong to the family of BHCs.

### Acknowledgments

We thank Dr. Carole Jordan for her careful reading of the manuscript and her helpful comments.